# Dielectric and energy-storage properties of $Ba_{0.85}Ca_{0.15}Zr_{0.10}Ti_{0.90}O_3$ ceramics with $BaO$–$Na_2O$–$Nb_2O_5$–$WO_3$–$P_2O_5$ glass addition

A. Ihyadn,[a*] S. Merselmiz,[a] D. Mezzane,[a,b] L. Bih,[c,d] A. Lahmar,[b] A. Alimoussa,[a]

M. Amjoud,[a] Igor A. Luk'yanchuk, [b,e] M. El Marssi,[b]

[a.] *IMAD-Lab, Cadi Ayyad University, Avenue Abdelkrim El Khattabi, P.B. 549, Marrakesh, 40000, Morocco.*

[b.] *Condensed Matter Physics Laboratory, University of Picardie Jules Verne, Amiens, France*

[c.] *Département Matériaux et Procédés, ENSAM Meknès, Université Moulay Ismail, Meknès, Maroc.*

[d..] *Equipe Physico-Chimie la Matière Condensée (PCMC), Faculté des Sciences de Meknès, Maroc.*

[e.] *Department of Building Materials, Kyiv National University of Construction and Architecture, Kyiv, 03680, Ukraine*

*Corresponding author: e-mail:ihyadn.abderrahim@gmail.com

**Abstract**

Lead-free $Ba_{0.85}Ca_{0.15}Zr_{0.10}Ti_{0.90}O_3$ (BCZT) ceramics with different $BaO$–$Na_2O$–$Nb_2O_5$–$WO_3$–$P_2O_5$ (BNNWP) glass content, forming $(1–x)$BCZT–$x$BNNWP lead free ceramics (abbreviated as BCZT$x$; $x$=0, 2, 4, 6, and 8wt%) were prepared by the conventional solid-state processing route. The effect of the BNNWP glass contents on the microstructure, phase structure, dielectric, ferroelectric, and energy storage properties of BCZT ceramics was investigated. The XRD analysis shows the coexistence of tetragonal and orthorhombic phases in BCZT pure. The SEM findings indicate that the average grain size decreases as the amount of BNNWP glass additives increases. In addition, BCZT ceramics modified with glass additions showed narrower hysteresis loops and a large electric field. The BCZT4 showed the highest recovered energy density of 0.52 J/cm$^3$ at 135kV/cm with an energy storage efficiency of 62.4%, which is increased by 6.6 compared to BCZT0 (0.075 J/cm3). The energy density was also calculated using the Landau-Ginzburg-Devonshire (LGD) theory.

# 1 Introduction

Pulse power technology has found significant application in electron beam, nuclear technologies, hybrid electric cars, and medical defibrillators. As a result of fast industrial development, it has continuously progressed into industrial and civil fields [1], [2]. Capacitors with higher storage density, faster charge–discharge rate, higher breakdown strength (BDS), and better thermal stability are required in modern electrical and electronic equipment [3], [4]. Among these requirements, low energy density is the primary barrier to limit their applications in high-power electronic devices [5].

The energy storage density is determined by the material's dielectric constant and the electrical breakdown resistance [6]. Among various energy storage materials, the dielectric ceramics exhibit relatively high dielectric constant and low breakdown strength (BDS) [3], [7].

BaTiO$_3$-based dielectric ceramics have been widely studied for energy storage devices due to their high permittivity and low energy loss [8]–[10]. Many recent studies have considered the effect of Ca$^{2+}$ and Zr$^{4+}$ replacing Ba$^{2+}$ and Ti$^{4+}$, respectively, in BaTiO$_3$ ceramics to form BZT–BCT solution. The results show that Ba$_{0.85}$Ca$_{0.15}$Zr$_{0.10}$Ti$_{0.90}$O$_3$(BCZT) ceramic exhibits excellent dielectric and piezoelectric properties due to the closeness to morphotropic phase boundary (MPB) [11]–[13]. Unfortunately, pure BCZT ceramic displays low dielectric strength and high-energy loss during charging and discharging processes, which results in poor energy storage properties and limits its use as an energy storage device. Hence, numerous researchers have developed several methods to improve the dielectric strength of ceramics in order to achieve sufficient energy storage performances [14], [15]. The findings reveal that these drawbacks can be overcome by adjusting the microstructure and the chemical composition of the ceramics. For instance, by adding some additives like oxides and glasses to the dielectric ceramics can considerably improve the dielectric breakdown strength (BDS) and hence the energy storage density [4], [7], [16].

Glass additives present an obvious advantage in improving the performance of dielectric ceramics. Appropriate amount of liquid phase will promote the rearrangement of ceramic particles, lowering sintering temperature and helping to remove the pores inside the green ceramic pellets as well as densifying ceramic powder compacts [4], [17]. Therefore, the glass content is crucial to enhance the energy storage density.

According to the recent literature studies, BCZT-based ceramics with glass additive showed enhanced dielectric strength and improved energy storage properties. Among various reported glass systems, B2O3–SiO2-based glasses, such as $B_2O_3$–$Al_2O_3$–$SiO_2$, $Bi_2O_3$–$B_2O_3$ $SiO_2$, and BaO–$B_2O_3$–ZnO present the particular interest [2], [18], [19]. The BDS of BCZT ceramics with glass additions were greatly improved by decreasing the grain size and densifying the microstructure [2], [18]. Nonetheless, a little number of researches was devoted to use phosphate glass with a low melting temperature compared to the $SiO_2$ and $B_2O_3$ glasses [20], [21]. The phosphate glasses have the supremacy for the advantage technology due to their simple composition combined with the strong glass forming character and low melting temperature [22].

In this work, $Ba_{0.85}Ca_{0.15}Zr_{0.10}Ti_{0.90}O_3$ (BCZT) ceramics with phosphate glass BaO–$Na_2O$–$Nb_2O_5$–$WO_3$-$P_2O_5$ (BNNWP) addition were synthesized using the conventional solid-state technique. The effect of BNNWP glass addition on dielectric and energy storage performances of BCZT ceramics was studied. The main objective is to obtain (1–x)BCZT–xBNNWP ceramics with enhanced energy storage properties. Furthermore, the energy storage densities were calculated using the Landau-Ginzburg-Devonshire (LGD) phenomenological theory. The modeling result confirms the experimental finding for BaTiO3-BaSnO3 as described by Yao et al. [23].

## 2 Experimental procedure

### 2.1 Synthesis of (1–x)BCZTx–BNNWP ceramics

A series of (1–x)BCZT–xBNNWP ceramics (x=0, 2, 4, 6, and 8wt%) designated as BCZT0, BCZT2, BCZT4, BCZT6, BCZT8 were prepared by the conventional solid-state reaction method. Meanwhile, BCZT powders and BaO–$Na_2O$–$Nb_2O_5$–$WO_3$–$P_2O_5$ glass were prepared in our previous works [13], [22]. The two powders were weighed according to the nominal composition of (1–x)BCZT–xBNNWP and then milled in an agate mortar with ethanol. Subsequently, the obtained mixtures were pressed into disks of 13 mm, and sintered in air at temperature ranging from 1350 to 1200 °C for 7h. The BCZTx ceramics were sintered at temperatures with the highest bulk density. The appropriate sintering temperature corresponds to the highest density and the value is 1350 °C, 1300 °C, 1275 °C, 1250 °C, and 1250 respectively when x=0, 2, 4, 6, and 8wt%.

### 2.2 Characterizations

The phase structure of BCZTx ceramics was analyzed by the X-ray diffraction (XRD, Panalytical™ X-Pert Pro spectrometer) using Cu-Kα radiation (λ~1.5406 Å). The density (d) of the BCZTx was measured at room temperature by Archimedes method. Scanning electron microscope (SEM, Tescan VEGA3) was used to examine the morphology of BCZTx ceramics. For electrical measurements, sintered ceramics coated with a silver paste form electrodes. The dielectric properties were measured by using an impedance analyzer (LCR meter hp 4284A 20Hz-1MHz). The polarization–electric field (P–E) hysteresis loops of a BCZTx ceramics with a thickness of 0.25mm were investigated with the CPE1701, PolyK, USA, with a high voltage power supply (Trek 609-6, USA).

## 3  Results and discussions

### 3.1  XRD analysis

The XRD patterns of BCZT$x$ ceramics are presented in the Figure 1. The prepared ceramic BCZT0 reveals the coexistence of orthorhombic (O) and tetragonal (T) phases. The existence of the quadratic phase is confirmed by the splitting of the peaks at 2$\theta$≈44-46° [13], [24]. Moreover, the formation of the BCZT sample at the Morphotropic Phase Boundary (MPB) with the coexistence of orthorhombic and tetragonal phases is evidenced by the presence of the triplet (022)$_O$/(200)$_T$/(200)$_O$ around 2$\theta$≈45° as reported in our previous work [13]. However, the coexistence of the O and T-phases disappear after the BNNWP glass addition to the BCZT ceramics. This phenomenon suggests that the structure of BCZT$x$ ceramics may transform from the coexistence of O and T-phases to the tetragonal phase. Figure 2 presents Rietveld fitted X-ray difraction patterns of (1-x)BCZT-xBNNWP (x=2wt% and 6wt%). For all compositions, the results reveal a significant compromise with the tetragonal structure (P4mm).  In addition, secondary phases (α) which could be attributed to the Ba$o$-Na$_2$O-Nb$_2$O$_5$-WO$_3$-P$_2$O$_5$ (BNNWP) glass's emerging phase are detected around 2$\theta$≈25°–35° when the glass is added. The corresponding lattice parameters and unit cell volumes of all samples are shown in Table 1. The evolution of the lattice parameters was clearly dependent on the BNNWP glass content. The unit cell volumes decreased with addition of BNNWP glass, which was consistent with the shift of the diffraction peaks to higher 2θ angles. Moreover, the result suggests that adding BNNWP glass led to the change in the lattice structure of the BCTZ samples. The change in phase structure may have an effect on the electrical properties of BCZT ceramics.

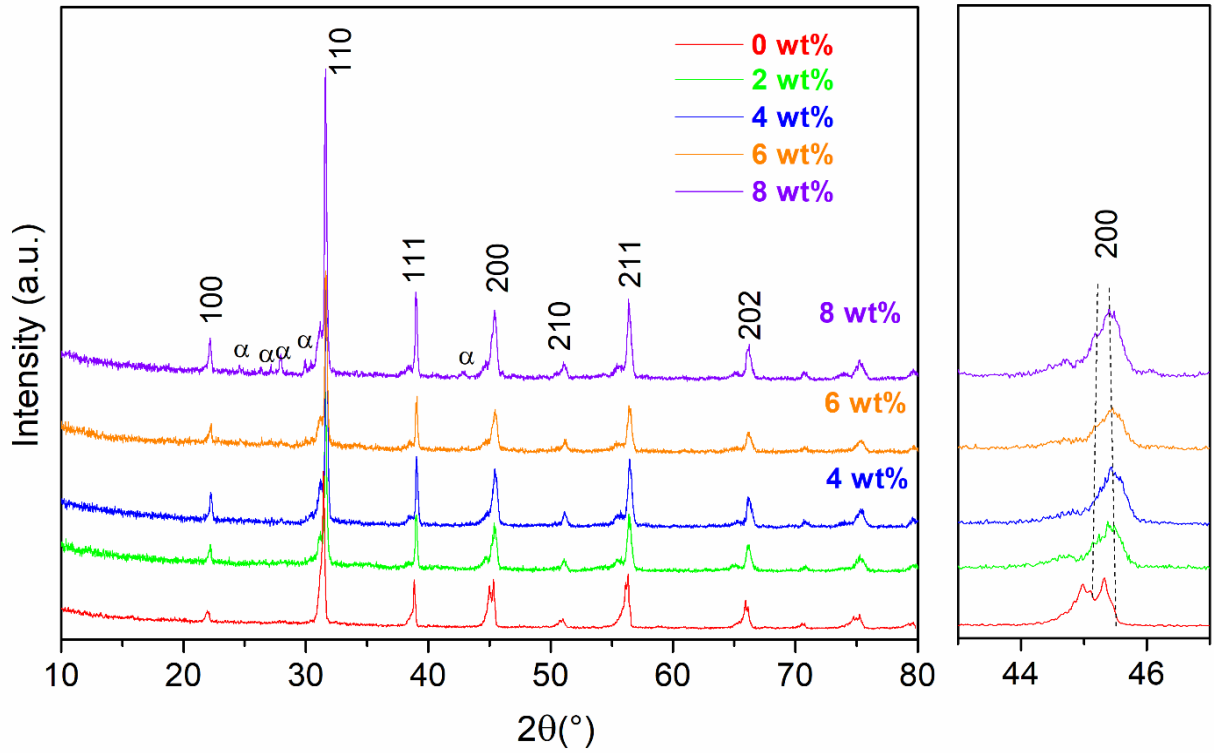

**Figure 1.** XRD patterns of BCZT*x* ceramics.

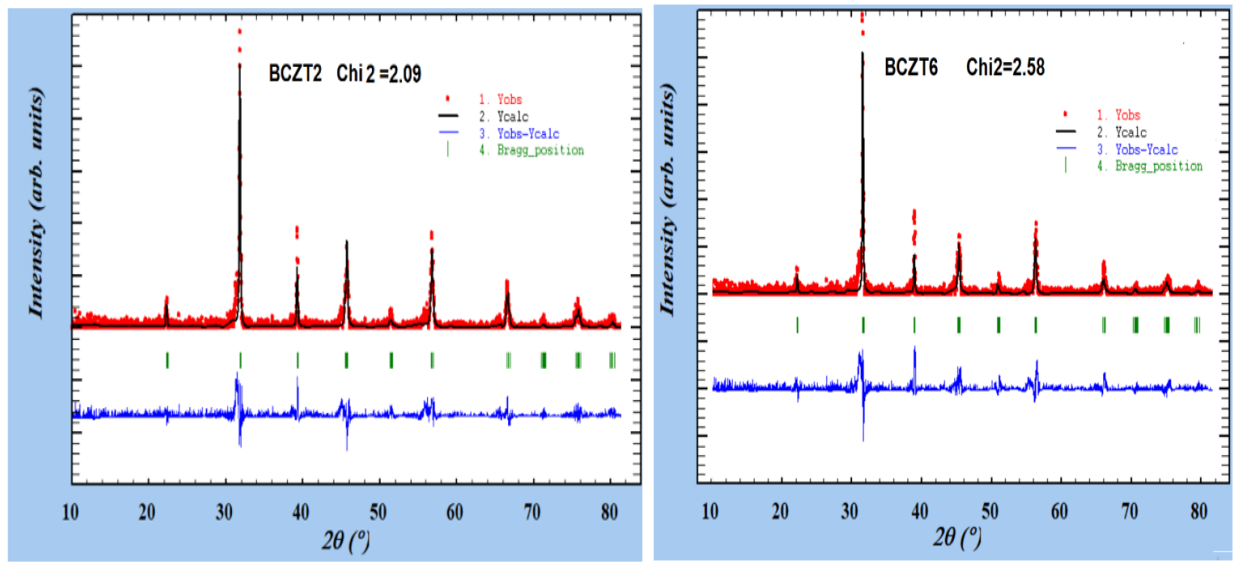

**Figure 2.** Rietveld ftted X-ray difraction patterns of (1-x)BCZT-xBNNWP for a x=2wt% and 6wt%.

**Table 1.** Lattice parameters and unit cell volumes of BCZTx ceramics.

| x(wt%) | a(Å) | b(Å) | c(Å) | V(Å$^3$) |
|---|---|---|---|---|
| **0** | 4.00403 | 4.00403 | 4.01404 | 64.35 |
| **2** | 3.96447 | 3.96447 | 3.98378 | 62.61 |
| **4** | 3.99529 | 3.99529 | 4.01605 | 64.10 |
| **6** | 3.99644 | 3.99644 | 4.01623 | 64.14 |
| **8** | 3.99646 | 3.99646 | 4.01853 | 64.18 |

**3.2  Microstructure analysis and density**

SEM micrographs of BCZT$x$ ceramics are shown in Fig. 3. The inclusion of small amounts of BNNWP glass reduces the average grain size of BCZT ceramics substantially. The average grain size decreases from 6.4 µm to 1.25 µm when the value of $x$ increases from 0 to 8wt%. This decrease may be due to the addition of the glassy BNNWP phase, which acts as an inhibitor of grain growth due to its low diffusion rate and low melting point. Moreover, the liquid glass phase existing between the grains can limit the migration of the grain boundary and prevent grain growth [2], [17]. The fine grain size and high density would contribute to the enhanced breakdown strength, which is a key element for high energy storage density [4], [7]. Moreover, the introduction of glass promotes the densification of BCZT$x$ ceramics, as shown in Fig. 4. The molten BNNWP glass at high temperature decreases the sintering temperature and improves the densification of BCZT$_x$ ceramics.

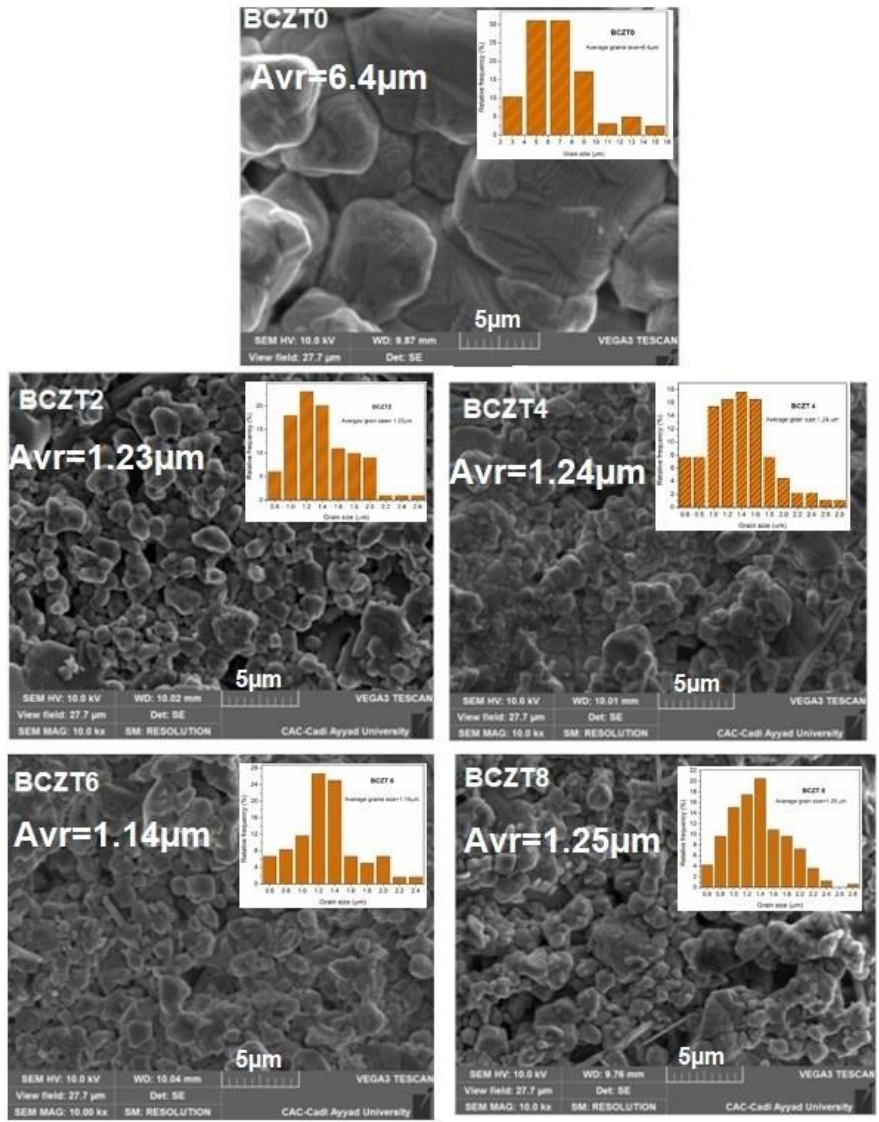

**Figure 3:** SEM images of BCZT*x* ceramics.

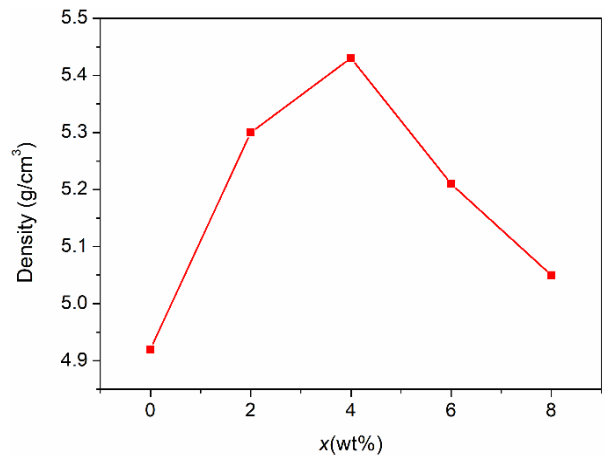

**Figure 4.** Density of BCZTx ceramics.

## 3.3 Dielectric properties

The temperature dependence of the dielectric constant ($\varepsilon_r$) of BCZT$x$ ceramics at different frequencies is shown in Fig.5. It can be seen that BCZT$x$ ceramics present two obvious polymorphic phase transitions corresponding to the orthorhombic-tetragonal (O–T) and tetragonal-cubic (T–C) transitions, respectively. The dielectric peak temperature of the O–T phase transition ($T_{O-T}$) is about 35.0, 24.7, 22.3, 23.4, and 29.5°C for BCZT0, BCZT2, BCZT4, BCZT6 and BCZT8 ceramics, respectively. In addition, the Curie temperature ($T_c$) decreases significantly when the glass content increases. The $T_c$ drops from 92 to 63 and then increases to 86, 84, 80°C when the glass additive content increases from 0 to 2, 4, 6 and 8 wt%, respectively. The decrease in the $T_c$ could be attributed to (i) the transition from a long-term order to a short-term order [25], which promotes the formation of polar nanoregions (PNRs) [26] and (ii) the internal clamping caused by the presence of immobile non-ferroelectric glass phase [27]. On the other hand, the value of the $\varepsilon_r$ at $T_c$ decreases significantly from 5400 to 524 with increasing additives in BNNWP glass. Hence, introducing of the BNNWP glass phase improves the thermal stability of the dielectric constant which could promote the energy storage performance of capacitors [28]. The temperature dependence of the dielectric constant($\varepsilon_r$) of BCZT$x$ ceramics at 1kHz is shown in Fig. 6. The Dielectric constant, strongly influenced by the addition of glass addition, decreases gradually with glass content. This dielectric change is resulted from the small ceramics grain size. It's reported that, smaller ceramic grain size generally leads to lower dielectric constant and the dilution of the low $\varepsilon_r$ of BNNWP glass ($\varepsilon_r$=65) contribute to the reduction of dielectric constant [22], [29]. Table 2 provides the dielectric properties obtained for all the ceramics.

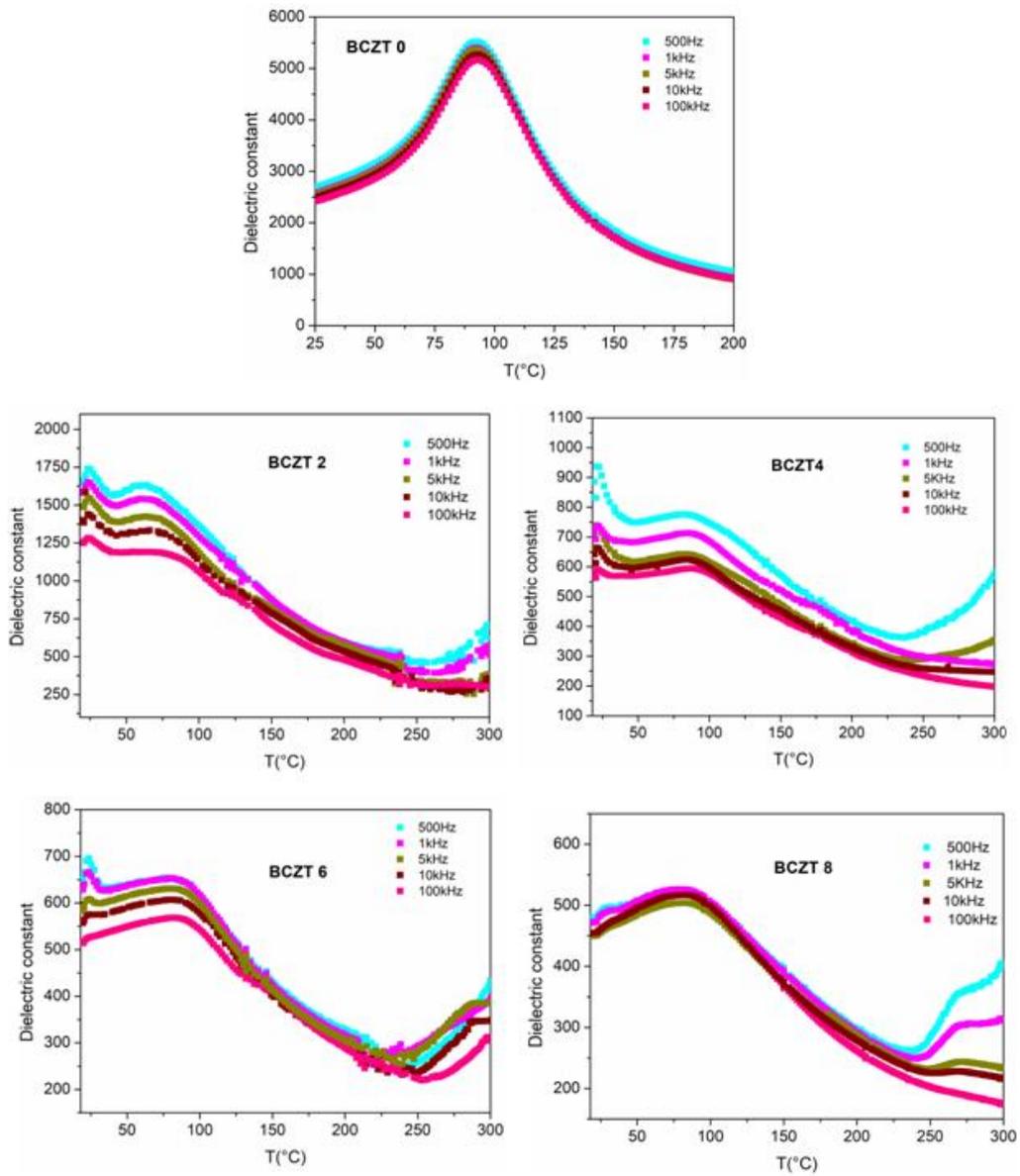

**Figure 5.** (a-e) Temperature-dependence of εr of BCZTxceramics at the frequency window of 500Hz-100kHz, (f) The evolution of Tc versus the glass content.

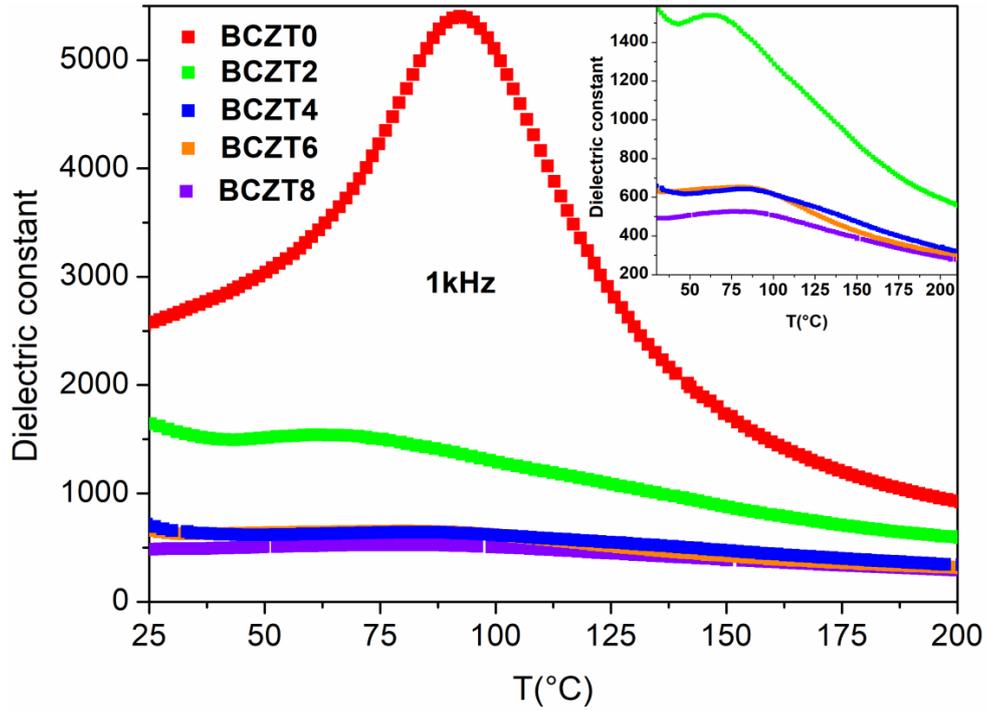

**Figure 6.** Temperature-dependence of $\varepsilon_r$ of BCZT$x$ ceramics at 1kHz.

**Table 2.** The dielectric constant and the peak temperatures of the phase transitions of BCZT$x$ ceramics.

| Sample | $T_{O-T}$ (°C) | $\varepsilon_r$ (30°C, 1kHz) | $T_c$ (°C) | $\varepsilon_m$ (1kHz) |
|---|---|---|---|---|
| **BCZT0** | 35 | 2640 | 92 | 5400 |
| **BCZT2** | 24.7 | 1587 | 63 | 1542 |
| **BCZT4** | 22.3 | 692 | 86 | 710 |
| **BCZT6** | 23 | 630 | 84 | 651 |
| **BCZT8** | 29.5 | 488 | 80 | 524 |

The temperature-dependence of the dielectric losses (tan δ) of BCZT$x$ ceramics at 1kHz is shown in Fig. 7. A significant increase in $tg\delta$ is observed near the O–T phase transition. This could be associated with the increase in conductivity, the internal stress and space charge induced by the interface between BCZT and glass. Note that except for the BCZT2 composition, a small increase of *tan δ* is observed above 40°C. However, the values of the dielectric losses are still low (<0.15).

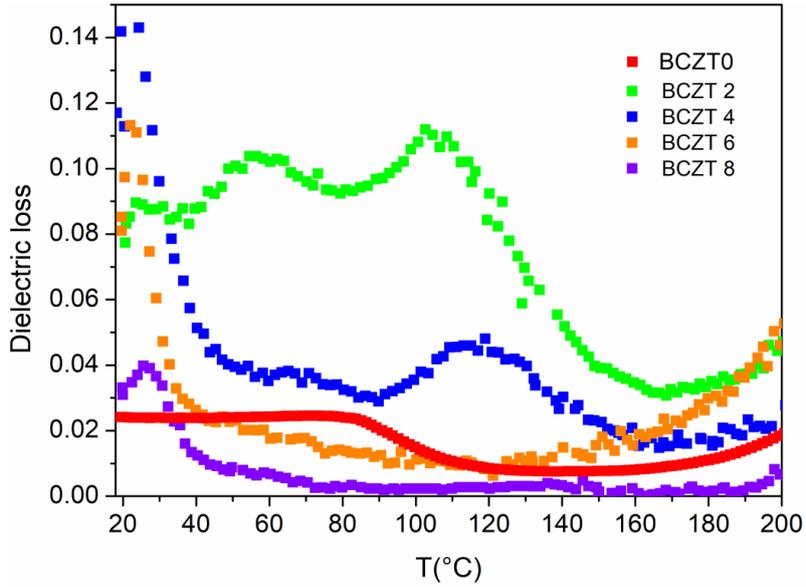

**Figure 7.** Temperature-dependence of *tanδ* of BCZT ceramics at 1kHz.

Fig.8 shows the frequency dependence of $ε_r$ and *tgδ* of BCZT*x* ceramics at room temperature. The frequency evolution of $ε_r$ reveals excellent dielectric stability for all compositions except BCZT0 and BCZT8. Furthermore, *tan δ* decreases with increasing the frequency in the range of 100Hz-10kHz. However, the evolution of *tgδ* remains stable in the frequency range of 10kHz-1MHz. This is beneficial for ceramics used in electrical energy storage [18], [28].

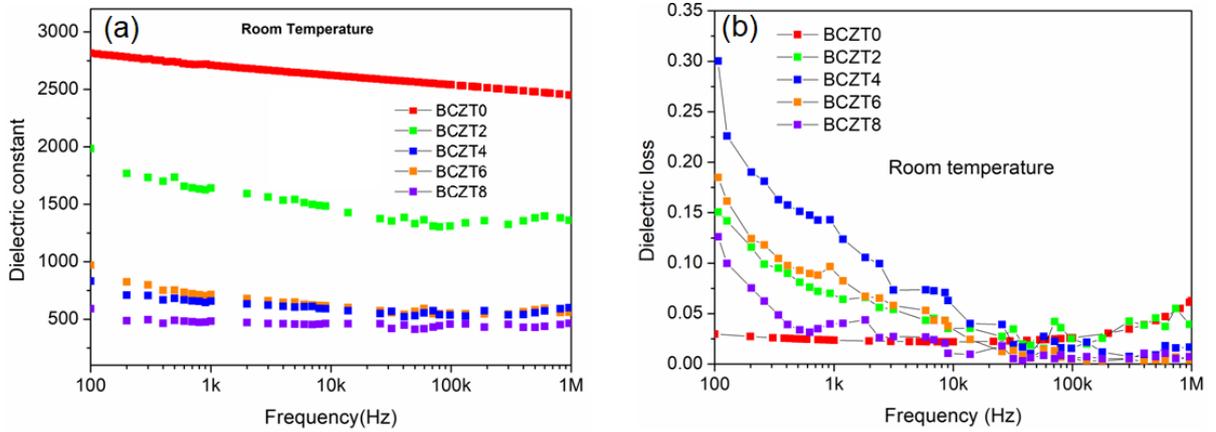

**Figure 8.** Room-temperature frequency-dependence of (a) $ε_r$ and (b) *tan δ* of BCZT*x* ceramics.

### 3.4 Energy storage performances

Fig. 9 (a-e) shows the room-temperature electric field-dependance of the *P–E* hysteresis loops of BCZT*x* ceramics at 100 Hz. It is evident that BCZT0 ceramic displays typical ferroelectric behavior with high maximal polarization ($P_{max}$) and remnant polarization ($P_r$) of 15 and 6 μC/cm², respectively, at 25 kV/cm. Obviously, the hysteresis loops become slimmer and the applied electric field increases as the glass content increases compared to BCZT0. At the same electric field, the polarization of the modified BCZT ceramics is lower than that BCZT0 because of the low permittivity of BNNWP glass additives. Fig. 9f shows the variation of the remnant polarization ($P_r$), the maximum polarization ($P_{max}$) and the maximal electric field ($E_{max}$) for all BCZT$x$ ceramics. It should be noted that the $P_{max}$ and $E_{max}$ values of BCZT ceramics increase until to reach a maximum and then decrease as the glass content increases. Besides, $P_r$ value decreases continuously with increasing the glass content. The highest $E_{max}$ of 135 kV/cm is obtained for the 4wt% sample. It is worthy to mention that the remnant polarization of the ceramics is consistent with the variation of the dielectric constant (Fig. 6). Despite the deterioration of the dielectric properties with further adding the glass, the decrease in $P_r$ is crucial to improve the energy storage density. These results could be attributed to the formation of polar nanoregions (PNRs) as observed in the behavior of the dielectric properties. Compared to normal ferroelectric domains, PNRs exhibit high dynamics, low energy barriers and good thermal stability, contributing to narrow P–E hysteresis loops with a low $P_r$ and a high breakdown strength [26].

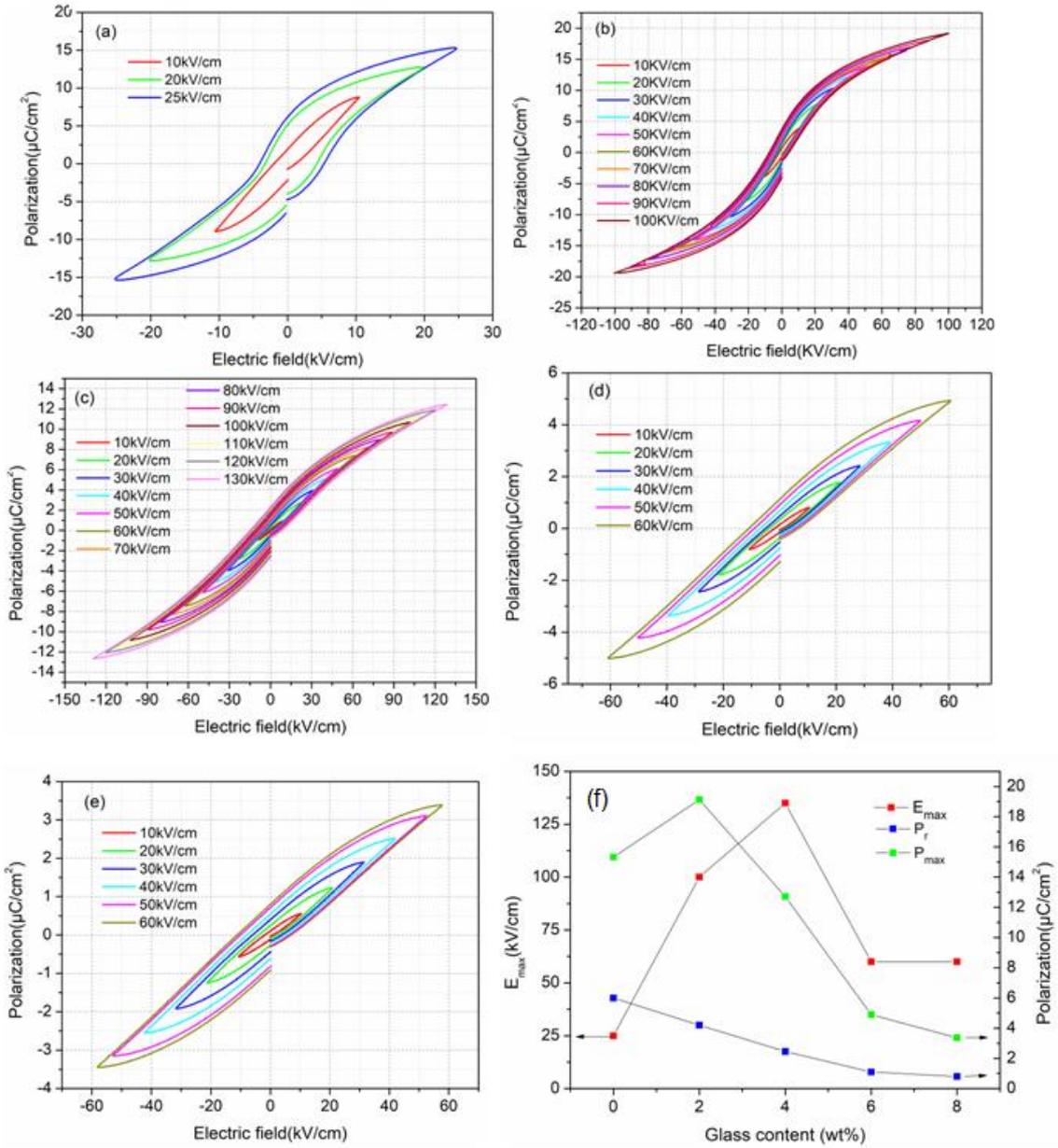

**Figure 9.** electric field-dependance of (a)–(e) the *P–E* hysteresis loops of BCZT*x* ceramics and(f) the *P*$_{max}$, *P*$_r$, and *E*$_{max}$ values of BCZT ceramics with glass additives.

The energy storage performances of BCZT*x* ceramics were determined from the recorded *P–E* hysteresis loops, where the total energy density ($W_{tot}$), the recovered energy density ($W_{rec}$, blue-colored areas) and the loss energy density ($W_{loss}$, red-colored areas) can be estimated using Eqs. (1) and (2) as illustrated in Fig. 10. Consequently, the energy efficiency ($\eta$) can be determined using Eq. (3) [27].

$$W_{tot} = \int_0^{P_{max}} EdP, \qquad (1)$$

$$W_{rec} = \int_{P_r}^{P_{max}} E\,dP, \qquad (2)$$

$$\eta\ (\%) = \frac{W_{rec}}{W_{tot}} \times 100 = \frac{W_{rec}}{W_{rec} + W_{loss}} \times 100. \qquad (3)$$

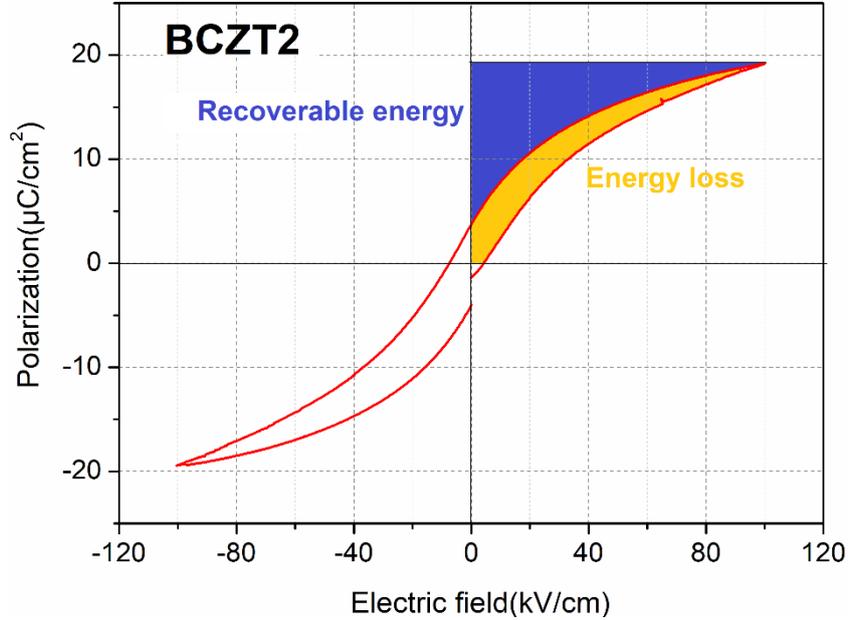

**Figure 10.** A schematic diagram of the energy storage densities estimated from the *P–E* hysteresis curve of BCZT2 sample.

Fig.11 shows the variation of, $W_{rec}$ and $\eta$ as a function of the applied electric field for all the samples. It can be seen that $W_{rec}$ increase with increasing the applied electric field. This indicates that the external voltage can obviously promote the recoverable energy density. However, it may have a negative impact on the energy efficiency, due to the increase of the energy loss resulting from internal relaxation polarization under high electric field. Indeed, pure BCZT ceramic exhibits a low energy efficiency ($\eta$<37%),Nevertheless by adding of glass, the energy efficiency ($\eta$) of BCZT*x* ceramics is highly improved and remains at a higher level due to the reduction of the energy loss. This result can be attributed to the decrease in Tc, which leads to the formation of polar nanoregions (PNRs), and the decrease of the $P_r$. Fig. 12 presents the variation of $W_{loss}$, $W_{rec}$ and $\eta$ as a function the glass content for all the samples at the maximum electric field. As expected, the recoverable energy density and the energy efficiency can be dramatically enhanced after the glass additions. Obviously, as the glass content increases, the recoverable energy density increases to reach a maximum of 0.52 J/cm³ in BCZT4, then decreases. Likewise, the maximal electric field value enhances substantially from 25 to 135 kV/cm as the glass content increases, then drops and remains constant. However,

BCZT6 and BCZT8 ceramics display a reasonably high maximal electric field (~62 kV/cm) compared to pure BCZT ceramic (~25 kV/cm). The addition of glass to BCZT ceramics significantly refined the microstructure, as evidenced by the SEM images. Previous studies reported that the improvement of *BDS* was due to the reduction of grain size, as the following relation, $E_{BDS} \propto G^{-b}$, where $E_{BDS}$ is the breakdown strength, *G* is the average grain size, and *b* is a constant [30]. Furthermore, the glass-forming phase at the grain boundary can also help to better improve the *BDS* as a thicker grain boundary can prevent the grains from breaking under high-applied voltage. The subsequent performance degradation may be caused by the coarsening of glass and forming discontinuity of grain boundary precipitates in BCZT6 and BCZT8 samples [31].

The optimum energy density is obtained for BCZT4 ceramic with the highest maximal electric field. It reaches 0.52 J/cm$^3$ at 135 kV/cm with an energy efficiency of ($\eta \sim 62.2\%$), which is 6.6 times higher than the pure BCZT ($W_{rec} \sim 0.075$ J/cm$^3$). As expected, the recoverable energy density and the energy efficiency can be significantly improved after the addition of glass to BCZT ceramics.

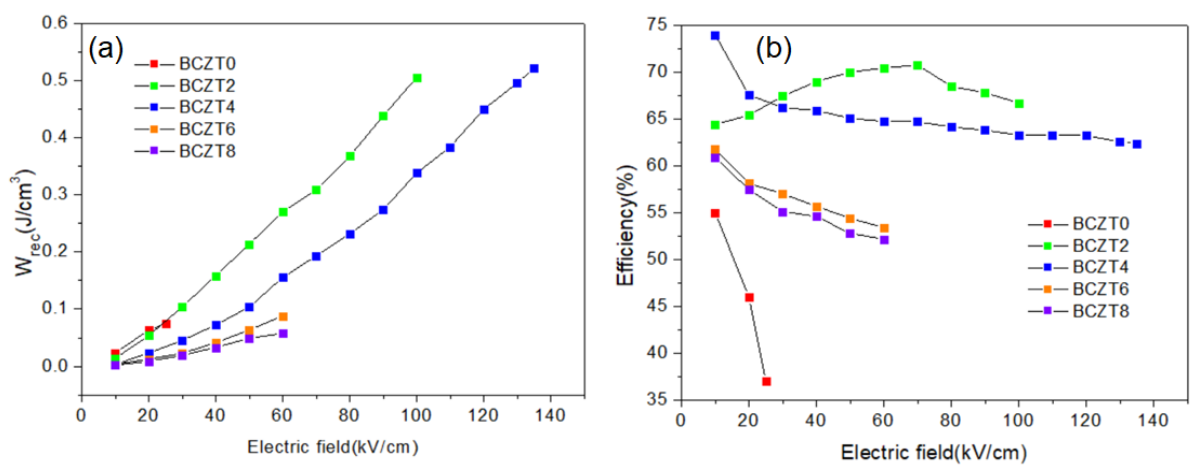

**Figure 11.** Electric field-dependance of (a) $W_{rec}$ and (b) $\eta$ for all the samples.

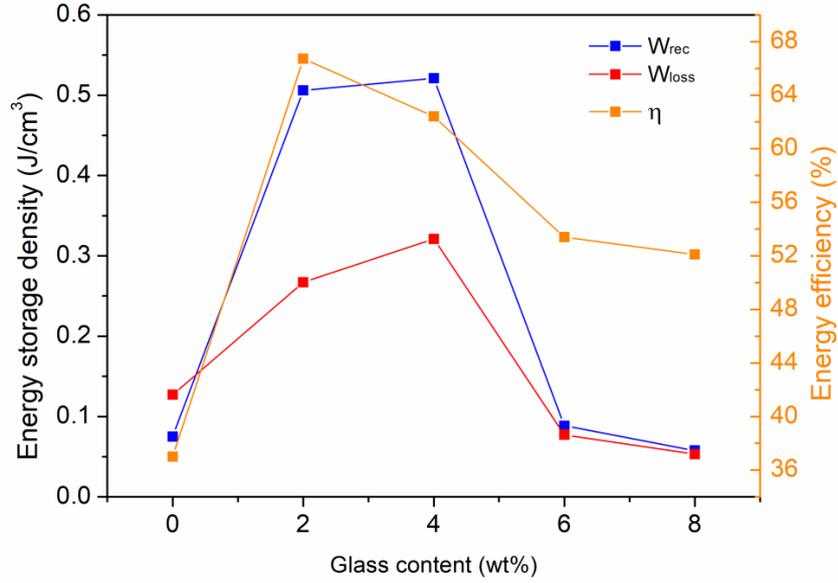

**Figure 12.** $W_{tot}$, $W_{rec}$ and $\eta$ of BCZT$x$ ceramics with different glass addition.

**Table 3.** Comparison of the dielectric constant and the energy storage properties of BCZT ceramic with other BT-based ceramics modified by glass addition reported in the literature.

| Samples | $\varepsilon_r$ (30 °C, 1kHz) | $E_{max}$ (kV/cm) | $W_{tot}$ (J/cm$^3$) | $W_{rec}$ (J/cm$^3$) | $\eta$ (%) | Ref |
|---|---|---|---|---|---|---|
| **BCZT2** | 1587 | 100 | 0.758 | 0.506 | 66.7 | This work |
| **BCZT4** | 692 | 135 | 0.834 | 0.521 | 62.4 | This work |
| Ba(Zr$_{0.2}$Ti$_{0.8}$)O$_3$-0.15(Ba$_{0.7}$Ca$_{0.3}$)TiO$_3$- BaO-SrO-TiO$_2$-Al$_2$O$_3$-SiO$_2$-BaF$_2$ | 500 | 96 | 0.192 | 0.149 | ~78 | [32] |
| Ba$_{0.95}$Sr$_{0.05}$Zr$_{0.2}$Ti$_{0.8}$O$_3$-MgO–CaO–Al$_2$O$_3$–SiO$_2$ | 1000 | 140 | 0.5 | 0.42 | 82.8 | [33] |
| Ba$_{0.4}$Sr$_{0.6}$Zr$_{0.15}$Ti$_{0.85}$O$_3$- SrO–B$_2$O$_3$–SiO$_2$ | 700 | 100 | 0.51 | 0.45 | 88.2 | [34] |
| Ba$_{0.95}$Ca$_{0.05}$Zr$_{0.3}$Ti$_{0.7}$O$_3$-MgO–CaO–Al$_2$O$_3$– SiO$_2$ | 1420 | 140 | 0.63 | 0..49 | 77.0 | [35] |
| Ba$_{0.85}$Ca$_{0.15}$Zr$_{0.1}$Ti$_{0.9}$O$_3$+B$_2$O$_3$-Al$_2$O$_3$-SiO$_2$ | 950 | 200 | 1.53 | 1.15 | 75 | [2] |
| Ba$_{0.85}$Ca$_{0.15}$Zr$_{0.1}$Ti$_{0.9}$O$_3$-Bi$_2$O$_3$–B$_2$O$_3$–SiO$_2$ | 895 | 330 | 2.34 | 2.12 | 90.5 | [18] |

To compare the energy storage performances of BCZT$x$ sample with other glass modified lead-free ferroelectric ceramics, Table 3 summarizes the $\varepsilon_r$, $W_{tot}$, $W_{rec}$ and $\eta$ of BCZT ceramics at the maximum electric field. At room temperature, BCZT4 sample shows enhanced energy storage density compared to other lead free ceramics [32], [33]. Note that, a high recoverable energy density of 0.192 J/cm$^3$ with an energy efficiency of 78% was obtained for Ba(Zr$_{0.2}$Ti$_{0.8}$)O$_3$-0.15(Ba$_{0.7}$ Ca$_{0.3}$)TiO$_3$ ceramic with 15 wt.% BaO-SrO-TiO$_2$-Al$_2$O$_3$-SiO$_2$-BaF$_2$ as glass-ceramic [32]. Meanwhile, Wang et al.[18] reported a large $W_{rec}$ of 2.12J/cm$^3$ and high $\eta$ of 90.5% in Ba$_{0.85}$Ca$_{0.15}$Zr$_{0.1}$Ti$_{0.9}$O$_3$/Bi$_2$O$_3$–B$_2$O$_3$–SiO$_2$ ceramic under a very high applied electric field of 330 kV/cm .

### 3.4. 1 Landau theory

More insight into the calculations of the total energy density parameters ($W_{rec}$, $W_{tot}$, $\eta$) can be obtained based the Landau-Ginzburg-Devonshire (LGD) phenomenological theory, describing the macroscopic phenomena in ferroelectric materials near the phase transition.

$$F = \frac{1}{2}aP^2 + \frac{1}{4}bP^4 - EP \quad (4),$$

where *a* and *b* are quadratic and quartic coefficients, respectively.

At equilibrium, one must have $\frac{\partial F}{\partial P} = 0$, which thus leads to:

$$E = aP + bP^3 \quad (5).$$

The electric field versus polarization (*E–P*) data for all considered temperatures can be fitted by Eq. (5), which shows that such equation is valid, but also allows to extract the *a* and *b* parameters for each sample. The examples of *a* and *b* coefficients are shown in Fig.13 for BCZT4 sample. These coefficients are important for energy storage of the energy density, according to the Landau model [36]: since they are involved in the expression

$$W = \int_0^{P_{max}} (aP + bP^3)dp = \frac{1}{2}aP_{max}^2 + \frac{1}{4}bP_{max}^4 . \quad (6)$$

Here $P_{max}$ is the polarization at $E_{max}$ and a and b are constant values. Eq. (6) therefore tells us that only three quantities completely govern the behaviors and values of the energy density, namely *a*, *b* and $P_{max}$. The Table 4 presents the parameter's *a* and *b* obtained and the energy density ($W_{rec}$ and $W_{tot}$) experimental and theoretical energy density calculated using Eq. (6).

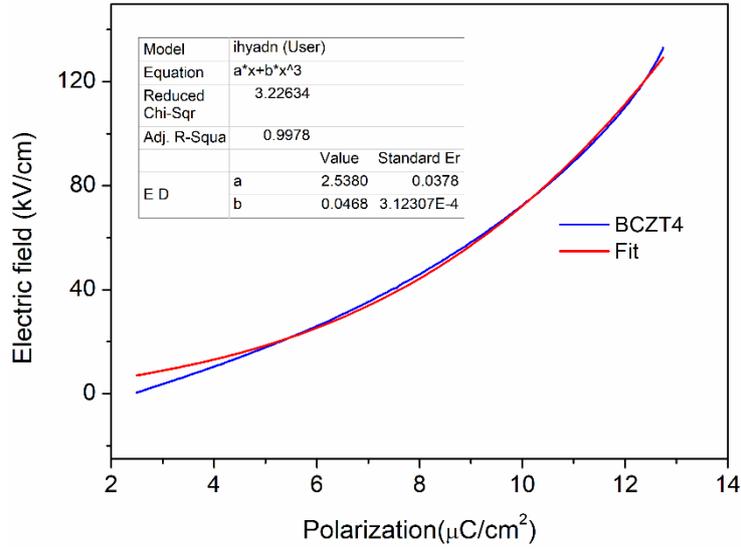

**Figure.13**.*E*–*P* hysteresis curves and the fitting parameters (*a* and *b*) during the discharge processes of BCZT4 sample.

**Table. 4**. The obtained parameters (*a* and *b*), the experimental and theoretical energy storage parameters ($W_{rec}$, $W_{tot}$ and $\eta$).

| Sample | $a_{(Wtot)}$ | $b_{(Wtot)}$ | $W_{tot(exp)}$ | $W_{tot(th)}$ | $W_{rec(exp)}$ | $W_{rec(th)}$ | $\eta_{exp}$ | $\eta_{exp}$ |
|---|---|---|---|---|---|---|---|---|
| BCZT0 | 1.3275 | 0.001448 | 0.202 | 0.173 | 0.75 | 0.70 | 37 | 40.3 |
| BCZT2 | 2.65793 | 0.00681 | 0.758 | 0.721 | 0.50 | 0.492 | 66.72 | 68.2 |
| BCZT4 | 8.95565 | 0.00878 | 0.803 | 0.782 | 0.52 | 0.512 | 62.4 | 65.47 |
| BCZT6 | 14.51299 | -0.11058 | 0.165 | 0.160 | 0.84 | 0.917 | 53.4 | 57.3 |
| BCZT8 | 20.78449 | -0.39079 | 0.110 | 0.106 | 0.577 | 0.587 | 52.14 | 55.42 |

Importantly, the theoretical energy density parameters ($W_{rec}$, $W_{tot}$ and $\eta$) calculated by the Landau theory are in agreement with the experimental results. Note that the coefficients *a* and *b* are temperature-dependent in relaxor ferroelectrics, which could be the main reason of the difference between the theoretical and the experimental results [37].

## 3.5  Conclusion

Based $Ba_{0.85}Ca_{0.15}Zr_{0.1}Ti_{0.9}O_3$ ceramics with different $BaO–Na_2O–Nb_2O_5–WO_3–P_2O_5$ glass content were prepared via the solid-state reaction method. The addition of glass results in a decreased average grain size and a denser microstructure. The dielectric properties showed that the dielectric constant decreased with increasing the BNNWP glass content. Fine hysteresis loops with higher electric field are observed in BCZT*x* ceramics with the glass additive. The maximal electric field of BCZT4 ceramic was remarkably improved for almost five times with respect to the pure BCZT ceramic. As a result, BCZT4 ceramic showed the highest recovered energy density of 0.52 J/cm$^3$ with an energy efficiency of 62.4% at 135kV/cm, which is 6.6 times better than the pure BCZT ceramic (~0.075 mJ/cm$^3$). The calculated from the Landau-Ginzburg-Devonshire (LGD) theory t energy densities are in agreement with the experimental findings. These results make BCZT$x$ ceramics a promising candidate for high-energy storage applications.